\documentclass[aps,prl,twocolumn,groupedaddress,longbibliography]{revtex4-1}
\usepackage{graphicx}
\usepackage{color}
\usepackage{amsmath}
\DeclareMathOperator{\Tr}{Tr}
\newcommand\pictcnq{picene-$\mathrm{F_4TCNQ}$ }
\newcommand\unit[1]{\,\mathrm{#1}}
\newcommand\fref[1]{FIG.~\ref{#1}}

\begin{document}

%\preprint{}

\title{Gate Controlled Molecular Switch Based on \pictcnq Charge-Transfer Material}

\author{Torsten Hahn}
\email[]{torsten.hahn@physik.tu-freiberg.de}
\author{Simon Liebing}
\author{Jens Kortus}
%\thanks{}
\affiliation{Institute of Theoretical Physics, TU Freiberg, Leipziger Str. 23, D-09596 Freiberg}

\date{\today}

\begin{abstract}

We show that the recently synthesized charge-transfer material \pictcnq can be used as a gate-voltage controlled molecular switch. 
The \pictcnq system is compared with the extensive characterized anthraquinone-based molecular system, which is known to exhibit large switching ratios due to quantum interference effects. In case of \pictcnq we find switching ratios larger by one order of magnitude. 
Further, our calculations reveal that the \pictcnq system resembles remarkable well the I-V characteristics of a classical diode. The reverse-bias current of this molecular diode can be increased two orders of magnitude by an external gate voltage. Based on density-functional theory calculations we show that the hybrid states formed by the \pictcnq system are playing the key role 
to determine the transport properties. We further conclude that the tuning of quantum transport properties through hybrid states is a general concept which opens a new route towards functional materials for molecular electronics.

% Based on the nonequilibrium Green's function method (NEGF) and density functional theory (DFT) calculations, we theoretically evaluate the electronic transport properties of this novel molecular system weakly coupled to Au(111) contacts. We especially take into account the effects of external gate voltages on the I-V characteristics of the model device. 

\end{abstract}

\pacs{}
%\keywords{}
\keywords{charge transfer complex, molecular switch, quantum transport, 
antraquinone}
\maketitle

%\section{Introduction \label{sec:Intro}}

% The design and the synthesis of new functional molecules is one of the major challenges in the field of molecular electronics. The prospected benefits of a technology based on molecules rather than conventional semiconductors in terms of pricing, energy efficiency and production simplicity are very promising. However recently also skeptical reviews about the progress and feasibility of this approach appeared \cite{Coskun2012}.
The basic idea of molecular electronics is that typical functionality needed to build integrated circuits has to be realized by single molecules.
This idea was already part of the pioneering theoretical description of a molecular diode given by Ratner and coworkers many years ago \cite{Aviram1974}. Since then it was repeatedly discussed \cite{Mujica2002,Metzger2008,Zhou2011} why it is not easy to actually build a satisfying molecular version of the classical semiconductor diode. In practice we face the problem that in nearly all molecular devices the I-V characteristics is rather symmetrical even when using a highly asymmetric molecule. As a result the attempts to manufacture molecular diodes or transistors resulted in rather poor rectification or switching ratios  compared to conventional semiconductor devices \cite{Kubatkin2003}. The criteria for achieving diode-like high rectification ratios in molecular junctions where already suggested by Ratner et al. \cite{Aviram1974}. Their proposal was that organic charge-transfer materials where two molecular subunits each carry an opposite charge should be suitable for the task. However, from this theoretical work it was also concluded that strong coupling between the molecular subunits may completely screen the desired effect.

Recently this idea was picked up by \cite{Tsuji2012a} and they presented rectification ratios $R = 2\dots 3$.
In this paper we present theoretical results on an the recently synthesized charge transfer material \pictcnq \cite{Mahns2014}. We will show that we can reach much higher rectification ratios in the range $R=20...50$ which is at least one order of magnitude higher than other values reported before. We compare our theoretical results to calculations on an anthraquinone based molecular switch. The anthraquinone system was chosen because it is well characterized and known to perform as an molecular switch \cite{VanDijk2006,Markussen2010,Seidel2013}.  Furthermore theoretical calculations with different contact materials and reliable measurements are available for comparison with our results  \cite{Zhao2011,Zhao2012,Guedon2012,Valkenier2014}.
We also investigate the dependence of the I-V characteristics on an external gate voltage. Apart from electrochemical gating \cite{Darwish2012} or redox active switching \cite{Zhao2012} this seems to be the most promising way to manufacture working active electronic devices.

\section*{Results and Discussion}

% figures should be put into the text as floats.
% Use the graphics or graphicx packages (distributed with LaTeX2e)
% and the \includegraphics macro defined in those packages.
% See the LaTeX Graphics Companion by Michel Goosens, Sebastian Rahtz,
% entire page. There is no need to do explicit centering.
%
%
The material under consideraion here is build from picene and fluorinated TCNQ molecules forming molecular crystals or films with an 1:1 composition which can be synthesized by co-evaporation \cite{Mahns2014}. The main building block in the crystal is a dimer like structure of picene and TCNQ with flat orientation with respect to each other. Therefore, we start our discussion with a short comparison of the electronic structure of the free \pictcnq-dimer and anthraquinone derivate
based on density functional theory (DFT) calculations.

\begin{figure}
  \includegraphics[width=0.99\columnwidth]{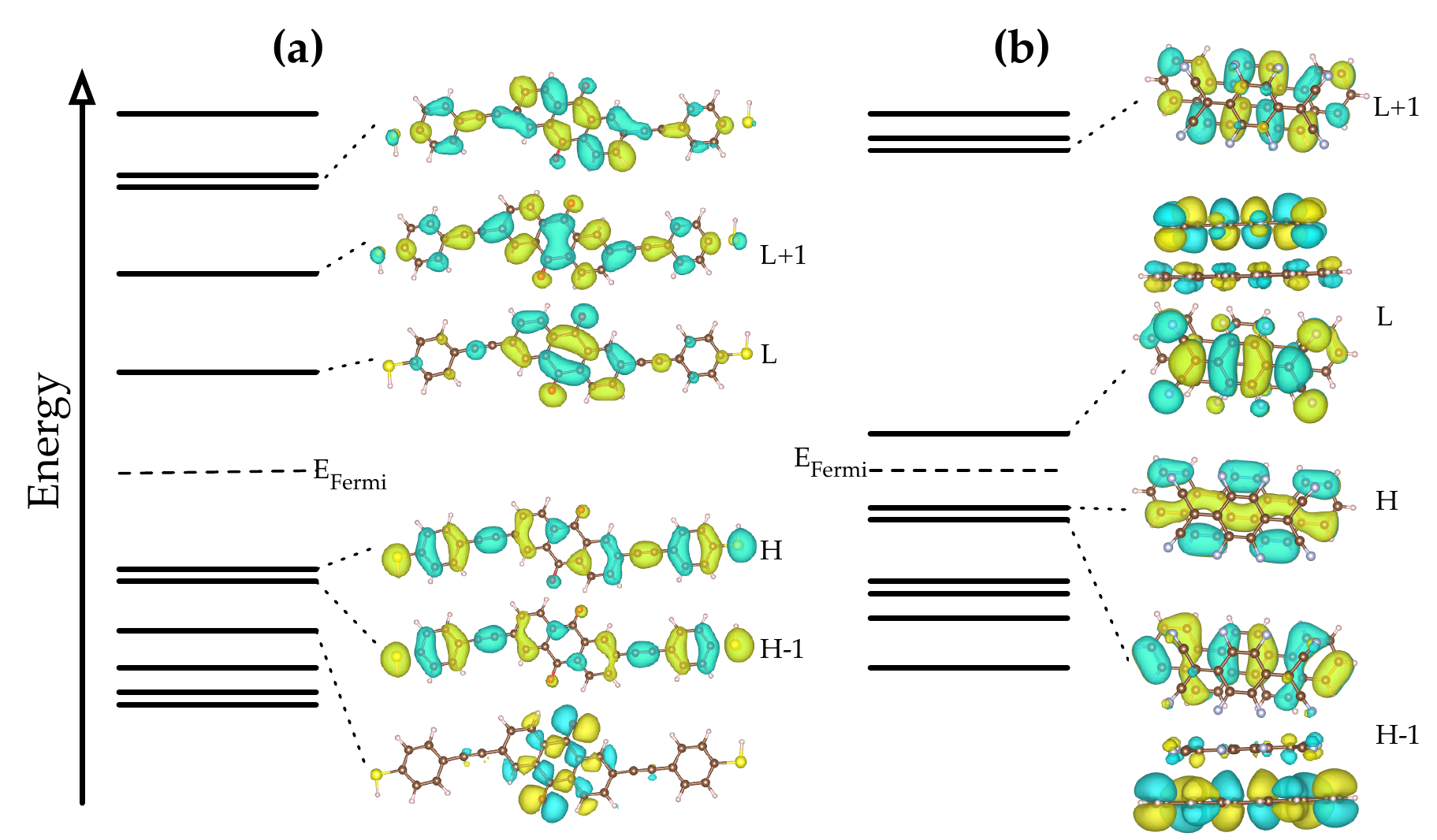}
  \caption{\label{fig:elevels} Electronic structure close to the Fermi level as obtained from density functional theory calculations for (a) the anthraquinone derivate and (b) the \pictcnq-dimer. }
\end{figure}

The electronic structure of the anthraquinone derivate as depicted in \fref{fig:elevels} a) has already been discussed in detail elsewhere \cite{Guedon2012,Seidel2013}. It shows the typical properties of a $\pi$-conjugated molecular material (semiconductor) having a HOMO - LUMO gap of about $1.6\unit{eV}$ and an almost symmetric level arrangement. The HOMO and HOMO-1 levels are delocalized and span the whole molecule including the thiol anchor units. The sprawling side structures guarantee that electron withdrawing or pushing effects of the anchor units modify only slightly the underlying electronic structure of the anthraquinone core.

\begin{figure*}
\includegraphics[width=0.99\textwidth]{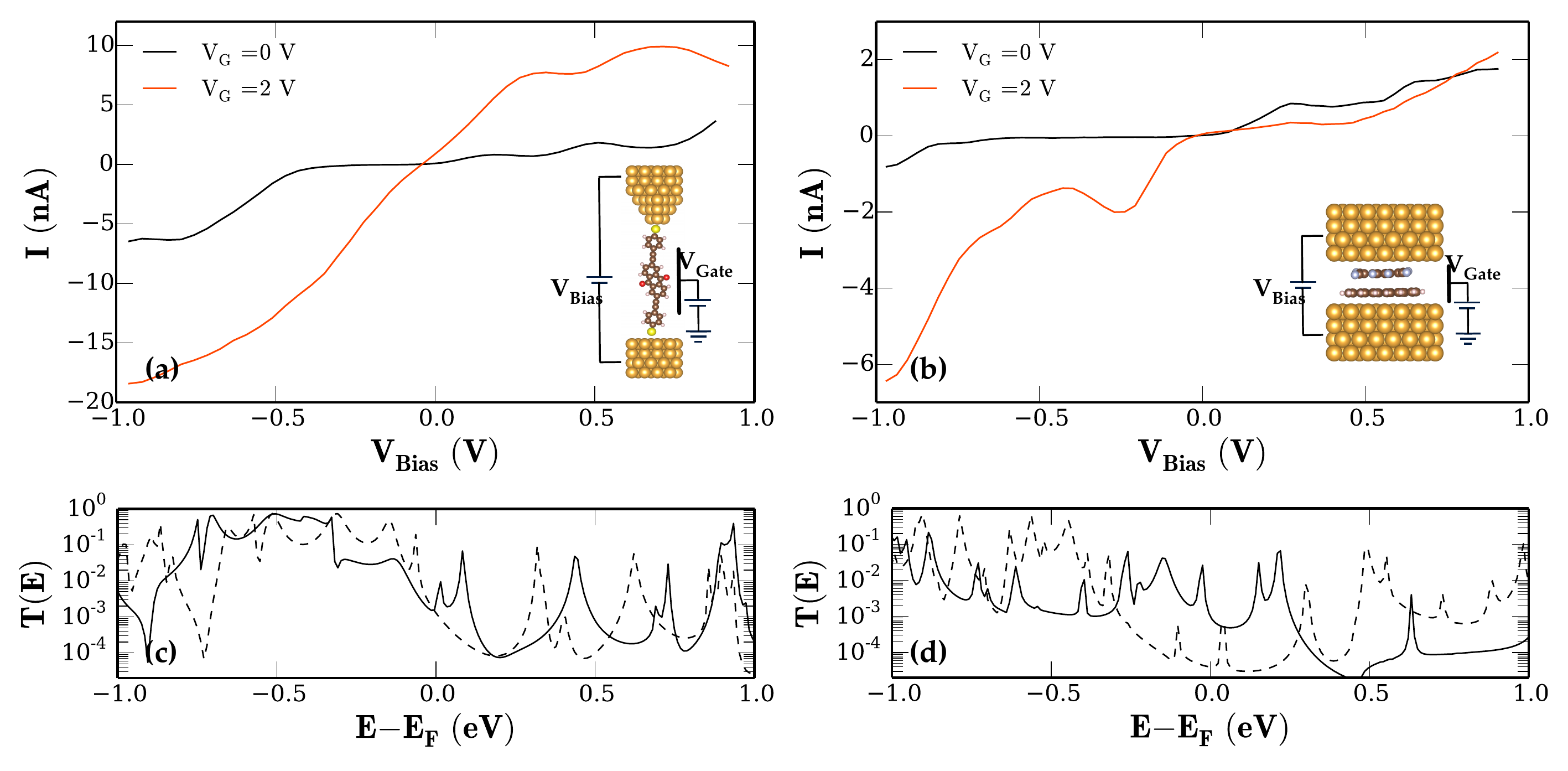}%
\caption{\label{negf_iv_curves}Calculated transport characteristics of the anthraquinone reference system (a) and the \pictcnq model device (b). The reference system shows moderate enhancement of the current in both bias directions. In contrast the \pictcnq shows a very strong increase only in reverse-bias region while the forward-bias current is nearly not affected by the external gate field. (c), (d) Respective transmission spectra $T(E)$ for the molecular junctions at zero (solid line) and $-0.75\unit{V}$ (dashed line) bias.}
\end{figure*}

The electronic structure of the \pictcnq system on the other hand is determined by the occurrence of hybrid orbitals which are formed between the $\pi$-conjugated picene and the $\mathrm{F_4TCNQ}$ acceptor. The HOMO-1 and LUMO orbital of \pictcnq are formed from states of the free picene and  $\mathrm{F_4TCNQ}$. This hybridization induces a charge transfer of about $0.2\unit{e}$ from the picene to the $\mathrm{F_4TCNQ}$ \cite{Mahns2014}.

Based on the electronic structure calculations the transport properties have been obtained using the non-equilibrium Green function formalism (NEGF).
Both molecular systems are sandwiched between two Au(111) leads.
\fref{negf_iv_curves} shows the I-V curves without gate voltage (black lines) for the a) anthraquinone and b) \pictcnq system. Both curves show typical details which are expected from the I-V-characteristics of a molecular junction. Due to the weak coupling of both molecular systems to the Au(111) leads both systems show features which can be attributed to distinct molecular orbitals. 

In a very simplified picture if bias voltage is rising then more orbitals will contribute to the conduction through the junction and the current increases. Peaks or regions of negative differential resistance (NDR) occur if distinct levels contribute to the conduction on low bias and do not contribute in the case of high bias voltages due to e.g.\ the lowering of the coupling strength between the molecular orbital and the lead \cite{Karthauser2011,Zhao2012}. Such a situation can be seen for example in the anthraquinone system at $\approx 0.2\unit{V}$ and $\approx 0.5\unit{V}$ bias voltage and for the \pictcnq system at $V_{bias}\approx 0.25\unit{V}$.

The effect of increasing bias voltage on the electron transmission spectra of the two molecular junctions is depicted in \fref{negf_iv_curves} c) and d) respectively. In the zero bias transmission function the \pictcnq-system (solid line) has a prominent feature at $0.2\unit{eV}$ above the Fermi level $(E_F)$ which can be attributed to the LUMO level of the dimer. The features right below $E_F$ are therefore linked to the HOMO, HOMO-1 etc. dimer orbitals. Above $0.3\unit{eV}$ \pictcnq exhibits no features relevant to transport. 

The main effects of an applied bias voltage to a molecular junction are (i) shifting the transmission spectra with respect to $V_{bias}$, (ii) strengthening and dampening of transmission features due to the bias induced changes in the molecule-lead coupling and (iii) widening of the energy window in which transmission peaks contribute to the current. Therefore if we apply a positive bias to the \pictcnq junction we are shifting transmission features which correspond to occupied HOMO-n orbitals into the energy window relevant for conduction. For negative bias the transmission peaks originating from the HOMO and LUMO are shifted out of that window and due to the large energy gap between the LUMO and LUMO+n levels (see \fref{fig:elevels}b) there are no additional levels which can  contribute to the conduction. 
Only in the case of very high bias voltages $> 1\unit{V}$ additional transmission features can appear and the current starts to rise also for negative bias.

For the anthraquinone junction the situation is completely different. We observe a somewhat lower density of transmission peaks for energies above $E_F$ (see \fref{negf_iv_curves}c). The changes in the bias voltage do not significantly change the overall density of transmission peaks in the energy window around $E_F$ contributing to the conduction. Hence the absolute value of the current through the junction is approximately the same whether we apply a negative or positive bias voltage. Our results are in good agreement with other theoretical estimates \cite{Zhao2011,Zhao2012}.
In the \pictcnq junction (\fref{negf_iv_curves}b) the current for the negative bias case stays almost at zero up to an $V_{bias}<-0.8\unit{V}$. For anthraquinone junction (\fref{negf_iv_curves}a) the current rises more or less symmetrical in case of forward and reverse bias.

\begin{figure*}
\includegraphics[width=0.99\textwidth]{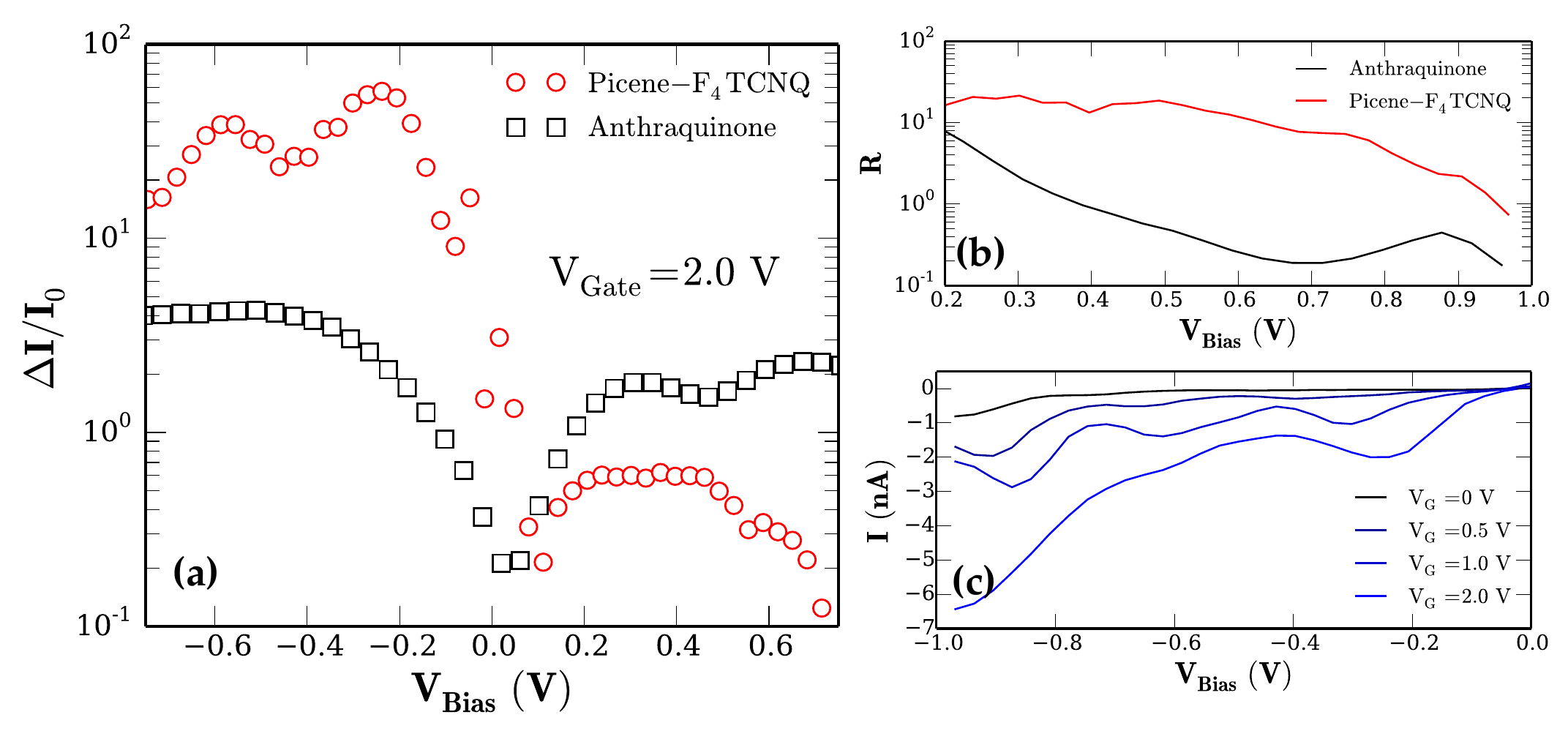}%
\caption{\label{negf_ampl}(a) Current across the \pictcnq model junction compared to the anthraquinone reference system as a function of the bias voltage at fixed $V_G=2\unit{V}$ ($I_0$ is the respective current without gate voltage). (b) Rectification ratio $R(V)=\left|I(V)/I(-V)\right|$ for the two molecular junctions under investigation. (c) Calculated change in the reverse bias current across the \pictcnq model junction as a function of the applied gate voltage $V_G$.}
\end{figure*}

This asymmetric behavior gives rise to large current rectification for the \pictcnq system. The rectification ratio can be estimated by $R(V)=\left|I(V)/I(-V)\right|$ \cite{Tsuji2012a}. We show a comparison of $R$ for the two systems in the voltage range of interest in \fref{negf_ampl}b. The highest $R$ are achievable for very low values of $V$ however for real applications voltages between $0.2 - 0.8\unit{V}$ seem to be more manageable whereas higher bias voltages may lead to rapid degradation in the organic material \cite{Schulze2008}. From our calculation we obtain rectification rations for $R_{Picene}\approx20$ which is much higher than the achievable maximum values for $R_{AQ}$.  Moreover for the anthraquinone system we find a steep decrease of the rectification up to $\approx0.6\unit{V}$ whereas the \pictcnq systems shows only a weak dependence in the considered bias range. With its large  rectification values the \pictcnq-system also outweighs already reported maximum values for other charge transfer and molecular materials by a at 
least one order of magnitude \cite{Tsuji2012a,Zhou2011}. The low variability of $R_{Picene}$ over a wide bias range seems more appropriate for real world applications. In consequence we propose the \pictcnq system as an molecular material to fabricate organic diodes due to its advantagoues forward and reverse bias properties.

Further, a required key element for implementing real molecular circuits is electrical switching. For that reason we compared the response of the two molecular systems to the application of an external gate voltage $V_{Gate}$ as schematically drawn in the insets of \fref{negf_iv_curves}. The I-V curves corresponding to a $V_{Gate}=2\unit{V}$ for the respective junctions are also shown as red lines in \fref{negf_iv_curves}a and b.
Both systems show a rather strong response to the application of an external gate field. For better quantification of the effect the change of the current due to the applied gate voltage $\Delta I=\left|I_0-I_{Gate}\right|$ normalized to current without gate $I_0$ for both systems is presented in \fref{negf_ampl}a.

For the anthraquinone junction we find a constant increase of the current for $\left|V_{bias}\right|>0.2\unit{V}$ between a factor two and three. Other authors report very large on/off ratios for anthraquinone junctions of $>1\times10^3$. However these ratios are achieved by chemical modification of the molecule itself and are potentially irreversible in contrast to the electrical switching presented here. Additionally the already mentioned theoretical \cite{Zhao2011,Zhao2012,Guedon2012,Valkenier2014} as well as recently measured data \cite{Guedon2012} on anthraquinone systems all show almost symmetrical behavior of the I-V characteristics.
The additional gate field in the junction induces two main effects. First the gate voltage shifts the molecular energy levels with respect to the energy window in which molecular orbitals contribute to the conduction (see \cite{Bogani2008}). In the present case the positive gate voltage results in  higher number of occupied molecular levels contributing and hence the current increases. The second effect of the gate field is the induction of changes of the electronic structure of the molecule itself. DFT calculations on the anthraquinone molecule with an applied electric field equal to the gate field in the transport calculations indicate that for example the HOMO-LUMO gap of the molecule is reduced and the molecular level alignment changes as well. In  \fref{negf_ampl}a one can see that the amplification of the current due to the gate field for the anthraquinone junction is almost symmetrical with respect to the bias voltage.

The same effects of a gate field also occur in case of the picene junction. However, due to the asymmetric character of the molecular orbitals around the Fermi level the impact on the I-V characteristics is much larger. Under forward bias the current is barely affected by the gate voltage. In fact the current shows even a small decrease. For the reverse bias case however we see a large increase of the current due to gate voltage. We achieve an maximum current increase by about $5\times10^2$ at $V_{bias}=0.3\unit{V}$. In \fref{negf_ampl}c we additionally show the dependence of the reverse bias current of the \pictcnq junction as a function of the applied gate voltage.

As mentioned before the large energy distance between the LUMO and LUMO+1 leads to almost no transmission for reverse bias. As the gate field shifts the molecular levels we see of course an current increase. However this level shifting alone does not explain the quantity of the observed effect. Here a further mechanism comes into play for the \pictcnq system. As discussed in the beginning the \pictcnq is a weakly bonded dimer with charge transfer of $\approx 0.2$ electrons from the picene to $\mathrm{F_4TCNQ}$. 
Our DFT calculations on the \pictcnq dimer with an electric field applied perpendicular to the stacking direction (= transport direction) reveal that the hybridization of the dimer itself depends strongly on the applied field. The applied electric field allows to tune the hybridization between the dimer components. This allows to lower the LUMO-LUMO+n distance drastically. The result is, that in the reverse bias case with the gate field switched on, the number of levels which account to the conduction is increased. This explains the very large switching ratios of $5\times10^2$.

In addition we wish to point out, that this behavior corresponds perfectly to the arguments given by Ratner and coworkers \cite{Aviram1974,Mujica2002} for achieving molecular rectification. Local weak links in a molecule, given in our case through the hybridization in the \pictcnq dimer, can result in large rectification ratios whereas strong bonding suppresses the effect. By direct modification of the hybrid levels using a gate field one can reach effective current switching. We believe that this mechanism is quite general for charge transfer systems and should 
be applicable to other dimer systems as well \cite{Lindner2012}.

\section{Summary} % (fold)

To summarize, the NEGF+DFT studies of the anthraquinone-Au(111) and \pictcnq-Au(111) systems have shown that the respective systems exhibit fundamentally different I-V characteristics. The anthraquinone system shows an approximately ohmic behavior for low bias voltages and the application of an external gate field results in an increase of the overall current through the junction which is almost symmetric for positive and negative bias voltages.

Due to hybrid dimer states close to the Fermi level the \pictcnq  I-V curve is very asymmetric with a pronounced diode-like forward/reverse current behavior. In contrast to the anthraquinone system the effect of an applied gate voltage is about two orders of magnitude larger in the reverse bias than in the forward bias case.

Further, we have shown that the anthraquinone system can also be seen as a electrically controllable switch. However, in terms of achieving maximum switching ratios the \pictcnq junction shows a clear benefit and can be seen as an molecular transistor in terms of classical circuit elements. The anthraquinone system on the other hand offers almost symmetric and linear I-V characteristics in the $V_{bias}\pm0.3\unit{V}$ range and may be better utilized as operational amplifier.

Consequently, we propose to use the pure organic interface between picene and $\mathrm{F_4TCNQ}$ as a straightforward way to manufacture a molecular switch with very large switching ratio or a molecular transistor / amplifier.
~\\

\begin{acknowledgments}
Financial support by the Deutsche Forschungsgemeinschaft within the Forschergruppe FOR 1154, project KO1924/5 and by the saxonian cluster of excellence ``Structure Design of Novel High-Performance Materials via Atomic Design and Defect Engineering (ADDE)'' is gratefully acknowledged. We especially thank the ZIH Dresden for providing extensive computational resources and support.
\end{acknowledgments}

\section{Methods}
The ground state electronic structure of the molecules was investigated using the all-electron density functional theory (DFT) \textsc{NRLMOL} program package which achieves a high level of numerical accuracy (see \cite{Pederson2000a,Porezag1999} and references therein). For the exchange correlation GGA/PBE \cite{Perdew1996a} was used and in all calculations dispersion correction utilizing the DFT-D2 method \cite{Grimme2006} was included. The geometry of the molecules was optimized using a gradient approach, the relaxation was terminated once all atomic forces were below {0.05}{~eV/\,$\mathrm{\mathring{A}}$}. 
We applied the nonequilibrium Green's function method (NEGF) for self-consistent calculation of the electron transport properties as implemented in the \textsc{GPAW} code \cite{Chen2012,Enkovaara2010} to investigate the I-V characteristics of our model devices. For the transport calculations the electronic structure is obtained by DFT calculations using the common approach of constructing a model device where the molecule of interest together with additional electrode atoms (scattering region) are sandwiched between two semi-infinite (metallic) electrodes. In our case we used at least three additional Au(111) layers on each side of the molecule to construct the scattering region followed by a further geometry optimization step of the system where the topmost two gold layers together with the attached molecules were allowed to relax. For the scattering region as well as for the leads a localized double-$\zeta$ polarized basis set was used.  Schematic drawings of the used model junctions are shown in the insets if \fref{negf_iv_curves}. The whole system can be subject to an external bias and/or gate voltage. The electronic structure of the scattering region and therefore the I-V curves are calculated self-consistently in the presence of such external fields. To support the deductions in this paper we just repeat the key facts of the DFT-NEGF method of use whilst a detailed discussion of the method can be found in the cited literature.
The \textsc{GPAW} transport code uses the Green's function of the central region defined by
\begin{equation}
  G(E) = \left[ES-H_C-\Sigma_L(E)-\Sigma_R(E)\right]^{-1}
\end{equation}
where $S$ and $H_C$ are the overlap and Hamilton matrix of the scattering region written in the localized basis. $\Sigma_{L/R}$ are the respective self energies of the leads. After self-consistency in the cycle $G\rightarrow n(r)\rightarrow v(r)\rightarrow H_C\rightarrow G$ is reached the transmission function is calculated by
\begin{equation}
  T(E,V)=\Tr\left[G(E)\Gamma_L(E)G(E)^\dagger\Gamma_R(E)\right]
\end{equation}
with $\Gamma_{L/R}(E)=i\left(\Sigma_{L/R}(E)-\Sigma_{L/R}(E)^\dagger\right)$. Therefore $T(E,V)$ gives the transmission probability of an electron having an energy $E$ under an applied bias (and gate) voltage $V$. Further the current through the junction is obtained by
\begin{equation}
  I(V)=\frac{2e^2}{h}\int^{\mu_R}_{\mu_L}T(E,V)dE
\end{equation}
were the electronic chemical potentials $\mu_{L/R}$ are connected to the applied bias voltage via $V=(\mu_L-\mu_R)/e$ ($e$ elementary charge) \cite{Meir1992}. De facto the current is calculated by integrating the self-consistent transmission function within the bias-dependent energy window spanned by $\mu_{L/R}$.



\begin{thebibliography}{100}
  \bibitem{Aviram1974} Aviram, A. and Ratner, M. A. ``Molecular rectifiers.'' Chem. Phys. Lett. 29, 277–283 (1974).
  \bibitem{Mujica2002} Mujica, V., Ratner, M. A. and Nitzan, A. ``Molecular rectification: why is it so rare?'' Chem. Phys. 281, 147–150 (2002).
  \bibitem{Metzger2008} Metzger, R. M. ``Unimolecular electronics.'' J. Mater. Chem. 18, 4364 (2008).
  \bibitem{Zhou2011} Zhou, K.-G. et al. ``Can azulene-like molecules function as substitution-free molecular rectifiers?'' Phys. Chem. Chem. Phys. 13, 15882–90 (2011).
  \bibitem{Tsuji2012a} Tsuji, Y., Staykov, A. and Yoshizawa, K. ``Molecular Rectifier Based on π–π Stacked Charge Transfer Complex.'' J. Phys. Chem. C 116, 2575–2580 (2012).
  \bibitem{Mahns2014} Mahns, B. et al. ``Crystal Growth, Structure, and Transport Properties of the Charge-Transfer Salt Picene/2,3,5,6-Tetrafluoro-7,7,8,8-tetracyanoquinodimethane.'' Cryst. Growth and Design 14, 1338-1346 (2014).
  \bibitem{VanDijk2006} van Dijk, E. H., Myles, D. J. T., van der Veen, M. H. and Hummelen, J. C. ``Synthesis and properties of an anthraquinone-based redox switch for molecular electronics.'' Org. Lett. 8, 2333–6 (2006).
  \bibitem{Markussen2010} Markussen, T., Schiötz, J. and Thygesen, K. S. ``Electrochemical control of quantum interference in anthraquinone-based molecular switches.'' J. Chem. Phys. 132, 224104 (2010).
  \bibitem{Seidel2013} Seidel, N. et al. ``Synthesis and properties of new 9,10-anthraquinone derived compounds for molecular electronics.'' New J. Chem. 37, 601 (2013).
  \bibitem{Zhao2012} Zhao, P. and Liu, D.-S. ``Electronic Transport Properties of an anthraquinone-Based Molecular Switch with Carbon Nanotube Electrodes.'' Chinese Phys. Lett. 29, 047302 (2012).
  \bibitem{Zhao2011} Zhao, P. et al. ``First-principles study of the electronic transport properties of the anthraquinone-based molecular switch.'' Physica B: Condens. Matter 406, 895–898 (2011).
  \bibitem{Kubatkin2003} Kubatkin, S. et al. ``Single-electron transistor of a single organic molecule with access to several redox states.'' Nature 425, 698–701 (2003).
  \bibitem{Guedon2012} Guédon, C. M. et al. ``Observation of quantum interference in molecular charge transport.'' Nat. Nanotechnol. 7, 305–9 (2012).
  \bibitem{Valkenier2014} Valkenier, H. et al. ``Cross-conjugation and quantum interference: a general correlation?'' Phys. Chem. Chem. Phys. 16, 653–62 (2014).
  \bibitem{Darwish2012} Darwish, N. et al. ``Single Molecular Switches: Electrochemical Gating of a Single Anthraquinone-Based Norbornylogous Bridge Molecule.'' J. Phys. Chem. C 116, 21093–21097 (2012).
  \bibitem{Karthauser2011}Karthäuser, S. ``Control of molecule-based transport for future molecular devices.'' J. Phys. Condens. Matter 23, 013001 (2011).
  \bibitem{Schulze2008} Schulze, G. et al. ``Resonant Electron Heating and Molecular Phonon Cooling in Single C60 Junctions.'' Phys. Rev. Lett. 100, 136801 (2008).
  \bibitem{Bogani2008} Bogani, L. and Wernsdorfer, W. ``Molecular spintronics using single-molecule magnets.'' Nat. Mater. 7, 179–86 (2008).
  \bibitem{Lindner2012} Lindner, S., Knupfer, M., Friedrich, R., Hahn, T. and Kortus, J. ``Hybrid States and Charge Transfer at a Phthalocyanine Heterojunction: MnPc/F16CoPc.'' Phys. Rev. Lett. 109, 027601 (2012).
  \bibitem{Pederson2000a} Pederson, M. R., Porezag, D. V., Kortus, J. and Patton, D. C. ``Strategies for Massively Parallel Local-Orbital-Based Electronic Structure Methods.'' Phys. status solidi 217, 197–218 (2000).
  \bibitem{Porezag1999} Porezag, D. and Pederson, M. ``Optimization of Gaussian basis sets for density-functional calculations.'' Phys. Rev. A 60, 2840–2847 (1999).
  \bibitem{Perdew1996a} Perdew, J. P., Burke, K. and Ernzerhof, M. ``Generalized Gradient Approximation Made Simple.'' Phys. Rev. Lett. 77, 3865–3868 (1996).
  \bibitem{Grimme2006} Grimme, S. ``Semiempirical GGA-type density functional constructed with a long-range dispersion correction.'' J. Comput. Chem. 27, 1787–99 (2006).
  \bibitem{Chen2012} Chen, J., Thygesen, K. S. and Jacobsen, K. W. ``Ab initio nonequilibrium quantum transport and forces with the real-space projector augmented wave method.'' Phys. Rev. B 85, 155140 (2012).
  \bibitem{Enkovaara2010} Enkovaara, J. et al. ``Electronic structure calculations with GPAW: a real-space implementation of the projector augmented-wave method.'' J. Phys. Condens. Matter 22, 253202 (2010).
  \bibitem{Meir1992} Meir, Y. and Wingreen, N. ``Landauer formula for the current through an interacting electron region.'' Phys. Rev. Lett. 68, 2512–2515 (1992).
\end{thebibliography}
\end{document}